\documentclass[a4paper,11pt]{article}
\pdfoutput=1 

\usepackage{jcappub} 
\usepackage[T1]{fontenc} 

\usepackage[margin=2.5cm]{geometry}
\usepackage{amssymb,amsmath,epsfig,esint}
\usepackage{calrsfs}
\DeclareMathAlphabet{\pazocal}{OMS}{zplm}{m}{n}

\newcommand{\Lb}{\pazocal{L}}

\usepackage{sidecap}
\sidecaptionvpos{figure}{t}
\usepackage{CJKutf8}
\usepackage{setspace,lipsum}
\usepackage{graphicx}
\usepackage{subcaption}
\usepackage[english]{babel}
\usepackage{booktabs}
\usepackage{cellspace}
\usepackage{multicol}
\usepackage{multirow}
\usepackage{diagbox}
\usepackage{rotating}
\usepackage[utf8]{inputenc}
\usepackage{fourier} 
\usepackage{array}
\usepackage{makecell}
\usepackage[toc,page]{appendix}
\usepackage[toc,page]{appendix}

\usepackage{caption}
\usepackage{mathtools}
\usepackage{comment}
\usepackage{setspace}
\usepackage{color}
\begin{document}
\title{\boldmath Coupled Quintessence scalar field model in light of observational datasets}


\author[a]{Trupti Patil}
\author[b,c,1]{Ruchika}\note{Corresponding author.}
\author[d]{Sukanta Panda}


\affiliation[a,d]{Department of Physics, Indian Institute of Science Education and Research Bhopal,\\
Bhopal Bypass Road, Bhauri,
Bhopal - 462066, Madhya Pradesh, India}
\affiliation[b]{Department of Physics, Indian Institute of Technology Bombay,\\ Main Gate Road, Powai, Mumbai,
Maharashtra 400076, India}
\affiliation[c]{Physics Department and INFN, Universit\`a di Roma ``La Sapienza'',\\ Ple Aldo Moro 2, 00185, Rome, Italy}


\emailAdd{trupti19@iiserb.ac.in}
\emailAdd{ruchika.ruchika@roma1.infn.it}
\emailAdd{sukanta@iiserb.ac.in}

\abstract{We do a detailed analysis of a well-theoretically motivated interacting dark energy scalar field model with a time-varying interaction term. Using current cosmological datasets from CMB, BAO, Type Ia Supernova, $H(z)$ measurements from cosmic chronometers, angular diameter measurements from Megamasers, growth measurements, and local SH0ES measurements, we found that dark energy component may act differently than a cosmological constant at early times. The observational data also does not disfavor a small interaction between dark energy and dark matter at late times. 
When using all these datasets in combination, our value of $H_0$ agrees well with SH0ES results but in 2.5$\sigma$ tension with Planck results. We also did AIC and BIC analysis, and we found that the cosmological data prefer coupled quintessence model over $\Lambda$CDM, although the chi-square per number of degrees of freedom test prefers the latter.

}

\maketitle
\flushbottom

\section{Introduction}
\label{sec:intro}
The significant observational evidence about Universe's dynamics \cite{SupernovaCosmologyProject:1997zqe, SupernovaSearchTeam:1998fmf, SupernovaCosmologyProject:1998vns} hint at the accelerated expansion at present. The elementary candidate to reveal the nature of this new physics includes Einstein's cosmological constant $`\Lambda$' with the equation of state (EoS) $`\omega$' = $-1$ in the $\Lambda$CDM model. Though this model provides a good fit to the latest observations, it lacks solutions to other significant problems \cite{Peebles:2002gy} in cosmology, especially to the fine-tuning problem and cosmic coincidence problem, also known as famous \textit{Why now?} problem. Along with this, we still have no information about the nature of Dark Energy and Dark Matter. All we know is that our universe is composed of two types of perfect fluids, one of which is baryonic and responsible for the deceleration of the universe and the growth of structures. The other one, dominant at late times, has negative pressure and is responsible for the acceleration of the present universe. Both the components have different equation of state parameter $\omega = P / \rho $. The baryonic component is known to be like radiation-dominated, stiff matter, or the dust-like universe. However, the determination of the equation of state for dark energy ($\omega_{DE}$) is still one of the important unresolved questions in cosmology. Recent studies \cite{Riess2004665, astier, eisenstein2005detection, MacTavish_2006,Komatsu_2009} rules out the possibility of $\omega_{DE}<< -1$ and other studies, SDSS and WMAP, \cite{tegmark2004cosmological} and \cite{Hinshaw_2009} put bounds as $-$1.67 < $\omega_{DE}$ < $-$0.62  and
$-$1.33 < $\omega_{DE}$ < $-$0.79 at present respectively.
\par
Very recent results \cite{Risaliti:2018reu} and \cite{Riess_2022} very clearly rule out the possibility of rigid $\Lambda$ by $5 \sigma$. In a complete cosmology model-independent way, the authors \cite{Capozziello:2018jya} showed at present, the universe requires a phantom equation of state ($\omega_{DE}<-1$) to explain the present observational datasets.

Thus, probing the cosmologists to explore beyond the standard model for dark energy candidates, perhaps dynamical dark energy model with varying EoS ($\omega$), see \cite{Copeland:2006wr, Bull:2015stt, Perivolaropoulos:2021jda, Schoneberg:2021qvd}, and references therein. The simplest possibility consists of a light scalar field such as a quintessence field, varying slowly during the Hubble time and can drive the accelerated expansion \cite{Zlatev:1998tr,Chiba:1999ka,dePutter:2007ny,Gonzalez-Diaz:2000glv,Duary:2019dfu,Caldwell:2005tm}. 
Some studies also argue that the quintessence scalar field should couple (non-gravitationally) with the matter as provided by \cite{PhysRevLett.81.3067} unless some special symmetry inhibits the coupling. Such a coupling or interaction between Dark Energy (DE)/quintessence scalar field and Dark Matter (DM) can be characterized by a continuous energy flow or momenta between the two components. Many studies in the literature prescribe a purely phenomenological approach to deal with interacting cosmologies of dark sectors. Then, other classes of models are motivated from a field theory perspective and are well motivated. Numerous observations \cite{Barreiro:2010nb,10.1093/mnras/sty2780,An_2018,Yang:2017yme,Fay:2016yow,Yang:2014okp, Piloyan_2014, Li:2013bya, Yang:2018euj, DiValentino:2019ffd, DiValentino:2020kpf, Gomez-Valent:2022bku} are in support of the theory of dark sector interaction, constraining energy transfer from DM to DE or the opposite paved their way into existence. Cosmological observations indicate that the coupling impacts the universe's evolution \cite{PhysRevD.79.043526,Zimdahl:2001ar}. Effects of interaction on the CMB and linear matter power spectrum \cite{Linton:2017ged,Olivares:2006jr} (for detailed review \cite{Wang_2016}), the structure formation at different scales and times and halo mass function \cite{Valiviita:2008iv, PhysRevD.91.063530, Carbone:2013dna, Amendola:2002mp, PhysRevD.85.023503} and on the behavior of the cosmological parameters \cite{He_2008, Binder:2006yc, Patil:2022uco,  Patil:2022ejk} have been investigated in respective works. Based on these, 
coupled quintessence models were also studied theoretically in several other contexts \cite{vandeBruck:2016jgg, Koivisto:2009fb,Gomes:2015dhl, PhysRevD.88.123512,Koivisto:2009ew,Barros:2015evi} to provide a viable cosmological solution. Interacting Dark Energy Models have also been studied extensively to alleviate "Hubble Tension" - which is now confirmed to be more than 5$\sigma$ after SH0ES 2022 results \cite{Planck:2018vyg, SPT-3G:2021eoc, ACT:2020gnv,  Riess_2022}. Along with Hubble Tension is coupled the $S_8$ tension or $r_{drag}$ tension, i.e., the sound horizon at the drag epoch tension because BAO and CMB measure the product of $H_0 $ and $r_{drag}$ and not $r_{drag}$ alone\cite{Verde:2019ivm, Evslin:2017qdn}. Several works \cite{Kumar:2017dnp,Yang:2018euj,Yang:2018uae,DiValentino:2021izs,Okamatsu:2021jil,DiValentino:2020vvd,Salvatelli:2014zta,Valiviita:2015dfa,Nunes:2016dlj,Marachlian:2015cia,Wang_2016, Banerjee:2020xcn,Lee:2022cyh,Montani:2023xpd,Archidiacono:2022iuu, Koivisto:2013fta,Chowdhury:2023opo} tried to address these tensions in cosmology but could only solve one tension or the other by keeping the same number of parameters. The only solutions \cite{Knox:2019rjx,DiValentino:2020zio, DiValentino:2021izs,Sen:2021wld, Escamilla:2023shf,Adil:2023exv,mehdi22,mehdi21, mehdi19} to resolve all cosmological tensions at present are those which are extremely fine-tuned or have more parameters and therefore increasing the uncertainties in the parameter space.

\par
In the present work, we do a detailed and elaborate study of late-time cosmological dynamics. We mainly study the well-motivated dark matter and dark energy interaction case theoretically and try to understand how current observational data sets like CMB, Type Ia Supernovae, cosmic chronometers, BAO, Masers, growth rate data, and local $H_0$ measurement from SH0ES study influences the constraints on different cosmological parameters.

\par
The work is structured as follows. In the next section, we present the model framework for the coupled dark sector and discuss its dynamics in brief. Later, we work out the linear dark matter perturbation evolution equation in section \ref{perturbation part}. Section \ref{observational analysis} describes the observational data and methodology adopted in this work. Section \ref{results and discussion} presents the results and thorough discussion from the observational analysis. We conclude the findings in Section \ref{final remarks}.

\section{Coupled field model and the Background dynamics}
In our study, we assume that the geometry of the universe describing flat, isotropic, and homogeneous universe is given by Friedmann-Lema$\hat{i}$tre-Robertson-Walker (FLRW) metric and the line element for it is given by
%
\begin{equation}
\label{eqn:1}
    \text{d}s^2 = -\text{d}t^2 + a^2(t) \big(\text{d}x^2+\text{d}y^2+\text{d}z^2 \big).
\end{equation}
which when transformed into spherical coordinates reads as
\begin{equation}
\label{eqn:1a}
    \text{d}s^2 = g_{\mu \nu}\text{d}x^\mu \text{d}x^\nu = -\text{d}t^2 + a^2(t) \bigg(\frac{\text{d}r^2}{1-kr^2}+r^2\text{d}\theta^2+r^2 \sin^2 \theta \text{d}\phi^2 \bigg).
\end{equation}
where k = $-$1, 0, and 1 denotes open, flat, and spatially closed universe.
We consider a canonical scalar field $\phi$ (as dark energy component) and non-relativistic (cold) dark matter component with pressure $P_{dm}$ = 0 undergoing an interaction that leads to a coupled quintessence dark energy model. Under this assumption, the action for dark sector coupling can be set up as
\begin{equation}
\label{eqn:2}
    S =  \int \text{d}^4x \sqrt{-g}\bigg( \frac{R}{2}  - \frac{1}{2}  g^{\mu \nu}  \partial_{\mu}\phi \partial_{\nu}\phi - V(\phi) + \sum_{i} {\Lb}^{i}_m (\chi_i, \phi) \bigg).
\end{equation}
$\Lb_m$ is the matter Lagrangian for different matter field species, which also depends on the field $\phi$ through the coupling. Different matter species ($i$) may experience different couplings \cite{PhysRevD.62.043511,Amendola:2001rc}. However, radiation is assumed uncoupled to dark energy because of conformal invariance. We also neglect the baryons since coupling to the baryons
is strongly constrained by the local gravity measurements \cite{PhysRevD.43.3873}. Thus, our study assumes that only dark matter couples to the dark energy scalar field.\\  
By taking varying the action \eqref{eqn:2} w.r.t. the inverse metric, we get Einstein’s gravitational field equation as follows
\begin{equation}
\label{eqn:3}
    G_{\mu \nu} = T_{\mu\nu} \equiv T_{\mu \nu} ^{(\phi)} + T_{\mu \nu} ^{(m)}.
\end{equation}
where $T_{\mu \nu}$ is the sum of $T_{\mu \nu}^{(\phi)}$ (energy-momentum tensor of DE component) and $T_{\mu \nu}^{(m)}$ (energy-momentum tensor of matter component). Since we allow interaction between the dark species, they do not evolve independently but coupled. Furthermore, they satisfy the local conservation equation in the form given:

\begin{equation}
\label{eqn:4}
    -\nabla^\mu T_{\mu \nu} ^{(\phi)} = Q_{\nu}=  \nabla^\mu T_{\mu \nu} ^{(dm)},
\end{equation}
where $Q_{\nu}$ expresses the interaction between dark matter and dark energy. 
\begin{equation}
\label{eqn:5}
    Q_{\nu} = F_{,\phi} \rho_{dm} \nabla_{\nu}{\phi}, 
\end{equation}
in which $F_{,\phi} \equiv \partial F / \partial \phi$. $F(\phi)$ = $F_0$ exp($ \beta \phi)$ is the coupling strength. $\beta$ is a constant. The radiation component, on the other hand, being not interactive with dark species, is separately conserved and follows $\nabla^\mu T_{\mu \nu} ^{(r)} = 0$. 
Using the Einstein equations, Friedmann equations for the coupled case scenario can be expressed as
\begin{equation}
\label{eqn:7}
\begin{split}
3H^2=  \frac{\dot{\phi}^2}{2}+ V(\phi) + F(\phi) \rho_{dm} +\rho_{r}, \\[4pt]
3H^2 + 2\dot{H}=-\bigg(\frac{\dot{\phi}^2}{2}- V(\phi) + F(\phi) P_{dm} +P_{r}\bigg).
\end{split}
\end{equation}
From the conservation of energy-momentum tensor (\ref{eqn:5}) or equivalently from equation (\ref{eqn:7}), one can easily derive energy conservation or continuity equation for each component.
\begin{equation}
\label{eqn:8a}
\begin{split}
\dot\rho_{\phi} +3H \rho_{\phi} (1+\omega_{\phi}) = Q\\
\dot\rho^{\prime}_{dm} +3H \rho^{\prime}_{dm}  = -Q\\
\dot\rho_{r} +3H \rho_{r} (1+\omega_{r}) = 0
\end{split}
\end{equation}
\vspace{0.1cm}
where $H \equiv \frac{\dot{a}}{a}$, indicates the time evolution of the universe. The dot indicates derivative with respect to time $t$, and  $\rho^{\prime}_{dm} = F(\phi) \rho_{dm}$ denotes the energy density of dark matter. 
Corresponding time component of interaction term in Eq.{\eqref{eqn:5}} is given by $Q  = F_{,\phi} \rho_{dm} \dot{\phi}$.

To study the dynamics of cosmological scalar fields in the presence of background fluid, it is easier to work with dimensionless variables. In our case, we introduce
\begin{align}
\label{dimensionless variables}
    x &=\frac{{\phi^ \prime}}{\sqrt{6}},&
    y &=\frac{{\sqrt{V(\phi)}}}{\sqrt{3}H},&
    \Omega_{dm} &= \frac{{F(\phi) \rho_{dm} }}{3H^2},&
    \Omega_{r} &=\frac{\rho_{r}}{3H^2}.&
\end{align}
along with 
\begin{equation*}
        \lambda = -\frac{{V,_{\phi}}}{V(\phi)},\qquad  m = \frac{{F,_{\phi} }}{F(\phi)}.
\end{equation*}
where $`$prime' denotes derivative w.r.t. ln(a). And we define energy fraction for dark energy as
\begin{equation*}
\Omega_{\phi} = x^2 +y^2
\end{equation*}
and dark matter energy density as

\begin{equation}
\label{constraint eq}
\Omega_{dm} = 1- (x^2 +y^2) - \Omega_{r}
\end{equation}
satisfying $ \Omega_{\phi} + \Omega_{dm} + \Omega_{r}=1$. 
We introduce one another dimensionless quantity $`\gamma$' for describing dark energy equation of state $`\omega_{\phi}$' as:
\begin{equation}
\label{gamma def}
    \gamma = 1+\omega_{\phi} =\frac{2x^2}{x^2+y^2}.
\end{equation}
Also, total (effective) equation of state parameter $`\omega_{eff}$' becomes
\begin{equation}
\label{total eos}
    \omega_{eff}=x^2-y^2+\frac{\Omega_r}{3}
\end{equation}
The coupled dynamical set of equations can now be expressed as:
%
\begin{equation}
\label{autonomous system}
\begin{split}
\gamma^\prime = \big(2-\gamma \big) \sqrt{3\gamma \Omega_{\phi}}  \bigg(-\sqrt{3\gamma \Omega_{\phi}} +\lambda \Omega_{\phi} +  m  (1-\Omega_{\phi}- \Omega_{r}) \bigg),  \\[0.5pt]
\Omega_{\phi}^\prime =  3(1-\gamma)\Omega_{\phi} (1-\Omega_{\phi})  + \Omega_{\phi} \Omega_{r} + m \sqrt{3\gamma \Omega_{\phi}} (1-\Omega_{\phi} - \Omega_{r})  ,\\[0.5pt]
\Omega_{r}^\prime =  \Omega_{r}(\Omega_{r} - 1) + 3 \Omega_{r} \Omega_{\phi} (\gamma - 1)   ,\\[0.5pt]
\lambda' = - \sqrt{3\gamma \Omega_{\phi}} \big(\Gamma_{\lambda}-1 \big) \lambda^2, \\[0.5pt]
m'=  \sqrt{3\gamma \Omega_{\phi}} \big(\Gamma_{m}-1 \big) m^2. 
\end{split}
\end{equation}

where 
\begin{equation}
\label{eqn:14}
        \Gamma_\lambda =\frac{{V,_{\phi \phi} V}}{V,_\phi^2},\qquad \Gamma_m =\frac{{F,_{\phi \phi} F}}{F,_\phi^2}.
\end{equation}
which for potential $V(\phi) \propto e^{-\lambda \phi}$ ($\lambda$ = constant) and for coupling parameter $F(\phi)$ = $F_0$ exp($ \beta \phi)$ gives unity. To solve the above autonomous system set, we assume the initial value of $\omega_\phi$ to be near -1, i.e., $\gamma \simeq 0.0001$\cite{Caldwell:2005tm}. Other quantities, namely $\Omega_\phi$, $\Omega_r$, $\lambda$, and m are free parameters.

Now, we study the dynamics of the system (\ref{autonomous system}) with the stationary points which satisfy the constraint (\ref{constraint eq}). The physically viable critical points ($x_c$, $y_c$, $\Omega_{dm_c}$) are

\begin{align*}
\label{critical points}
    P_1 &= \big(0,0,0 \big),&
    P_2^{(\pm)} &= \big(\pm 1,0,0 \big),&
    P_3^{(\pm)} &=\Bigg(\frac{\lambda}{\sqrt{6}}, \pm \sqrt{1-\frac{\lambda^2}{6}}, 0 \Bigg),&
    P_4^{(\pm)} &= \Bigg( \frac{2 \sqrt{\frac{2}{3}} \lambda}{1+\lambda^2}, \pm \frac{2 \sqrt{1+\frac{\lambda^2}{3}} }{1+\lambda^2} , 0 \Bigg).&
\end{align*}
Point $P_1$ explains the radiation-dominated universe as seen from Eq.(\ref{constraint eq}) and takes the value of effective EoS parameter $\omega_{eff}$ as 1/3. Points $P_2^{(\pm)}$ describe the universe dominated by the kinetic part of the scalar field. Thus the solution corresponds to stiff fluid with the effective equation of state parameter $\omega_{eff}= \omega_{\phi}=1$. Moreover, stationary points $P_3^{(\pm)}$ describe an accelerated universe when $\mid \lambda \mid <\sqrt{2}$. A corresponding de Sitter solution occurs only when $\lambda=0$ making $\omega_{eff}= -1$, i.e., the field potential plays the part of the cosmological constant. Points $P_3^{(\pm)}$ are physically acceptable under the constraint $\mid \lambda \mid < \sqrt{6}$. However, points $P_4^{(\pm)}$ produces a dark energy dominated accelerated solution when $\mid \lambda \mid < (2\sqrt{2}-1)$. 

We now study the stability of the above stationary points. The following are the eigenvalues corresponding to the stationary points 
\begin{equation*}
\label{eigenvalues12}
    e(P_1) =\left \{2,-1, \frac{1}{2} \right \}, \qquad
    e(P_2^{(\pm)}) =\left \{3, \quad \frac{1}{2} ( 3 \pm \sqrt{6}m), \quad 3 \mp \sqrt{\frac{3}{2}}\lambda \right \},
\end{equation*}
\begin{equation*}
\label{eigenvalues34}
     e(P_3^{(\pm)}) =\left \{ \frac{\lambda^2}{2}-3, \quad \lambda^2 -3, \quad  \frac{1}{2}(\lambda^2 +m\lambda -3) \right \}, \quad e(P_4^{(\pm)}) =\left \{ \frac{1}{2}+ \frac{2(m\lambda-1)}{1+\lambda^2}, \quad \frac{n\mp \sqrt{s}}{2(1+\lambda^2)^2}  \right \},
\end{equation*}
\vspace{0.2cm}
where, $n = 9+\lambda^2-9\lambda^4-\lambda^6$ and $s = 225+144\lambda^2+174\lambda^4-16\lambda^6-15\lambda^8$.
\newline
From the eigenvalues, we find that the point $P_1$ and the point $P_2^{(\pm)}$ exhibit saddle nature. Moreover, for $m=\lambda=0$, point $P_2^{(\pm)}$ exhibits instability too. Point $P_3^{(\pm)}$ becomes a stable attractor for  $\lambda^2 < 3$ and $m < \frac{3}{\lambda} - \lambda$ and it is a repeller when $\lambda^2 > 6$ and $m > \frac{3}{\lambda} - \lambda$. Otherwise, the point shows saddle behaviour. Furthermore, point $P_4^{(\pm)}$ acts as an attractor for $m < \frac{1}{4\lambda}(3+  \lambda^2)$ and $n < \sqrt{s}$ and becomes unstable when $m > \frac{1}{4\lambda}(3+  \lambda^2)$ and $n > -\sqrt{s}$. Point $P_4^{(\pm)}$ can also describe an oscillatory solution if $`s$' value becomes negative.

With these results, we now continue to investigate the perturbation theory for this model and its implications when confronted with cosmological observations.
\section{Linear Perturbation Equation}
\label{perturbation part}
Since the coupling modifies the evolution of matter perturbations and the clustering properties of galaxies, to gain further insight, we now study the evolution of the matter perturbations and, thence, include the linear growth rate data ($f\sigma_8$) from large-scale structure (LSS) survey. We re-scale the term $``F(\phi)\rho_{dm}$'' as $\rho_c$. Given this, we define the perturbation variable as $\delta_c \equiv \frac{\delta \rho_{c}}{\rho_{c}}$, where the subscript $c$ stands for the (cold) dark matter component and $\delta \rho_{c}$ is the deviation from the background dark matter density. Mathematically, the equation governing the growth of perturbation (in the Newtonian gauge) at linear regime
is given as
\begin{equation}
\label{perturbation eq}
        \delta_{c}^{''} + \Big(2+\frac{H^{'}}{H} \Big) \delta_{c}^{'} - \frac{3}{2} \Omega_{dm} \delta_{c} = 0.
\end{equation}
The introduction of coupling to the dark matter then modify the growth of matter perturbation as 
\begin{equation}
\label{coupled perturbation eq1}
        \delta_{c}^{''} +  \delta_{c}^{'} \bigg(2+\frac{H^{'}}{H} - \frac{F,_{\phi}}{F(\phi)} \phi^{'} \bigg) - \frac{3}{2} \delta_{c} \bigg( \Omega_{dm} - \frac{1}{3} \phi^{'2} \bigg) \big(1+  \frac{2 {F,_{\phi} }^2}{F(\phi)^{2}} \big)  =0.
\end{equation}
which can also be written in terms of dimensionless variables as
%
\begin{equation}
\label{coupled perturbation eq2}
        \delta_{c}^{''} = - \bigg(\frac{1}{2}- \frac{1}{2} \Omega_{r} -\frac{3}{2}(\gamma-1)  \Omega_{\phi} - m \sqrt{3\gamma \Omega_{\phi}} \bigg) \delta_{c}^{'} + \frac{3}{2}\bigg( 1-\Omega_{\phi}(1+\gamma)-\Omega_{r}  \bigg) \big(1+ 2m^2 \big)    \delta_{c} .
\end{equation}
%
We write the modified Hubble normalized, in terms of N = $\ln{a}$ as
\begin{equation}
\label{hz(n=lna)}
        \frac{H}{H_0} = \Big[\Omega_{0,\phi} e^{-3N(1+\omega_{\phi})} + \Omega_{0,dm} e^{-3N} e^{-m \sqrt{3\gamma \Omega_{\phi}}N}   + \Omega_{0,r} e^{-4N}  \Big]^{\frac{1}{2}}.
\end{equation}

The dark matter perturbations and the Hubble evolution for the uncoupled case are reproduced by setting $m$ = 0. We consider the time evolution of the universe at $N_i$ = $-7$ to ensure the evolution around the matter-dominated epoch. Therefore, we use the initial conditions as $\phi(N_i)$ = $\phi^{'}(N_i)$ = 0. And for the matter density contrast we take $\delta_{c}(N_i) = \delta_{c}^{'}(N_i) = 10^{-3} $. The dynamical set of equations (Eq. \ref{autonomous system}) and the perturbation equation (Eq. \ref{coupled perturbation eq2}) are addressed with the cosmological observations to constrain the interacting scalar field quintessence model and its parameters.
\section{Observational data}
\label{observational analysis}
We devote this section to presenting and describing the main observational data implemented to constrain the model parameters and the statistical analysis results.
\begin{itemize}
    \item The \textbf{Cosmic Microwave Background} (CMB) data are an effective probe for observational analysis of the cosmological models. In this work, we work with the CMB distance priors on the acoustic scale $l_A$ 
    leading to the alteration of the peak spacing and on the shift parameter $`\mathrm{R}$' 
    affecting the heights of the peaks. We consider using the \textit{Planck} 2018 compressed likelihood TT, TE, EE + lowE obtained by Chen et al \cite{Chen:2018dbv} (see Ade et al. \cite{Planck:2015bue} for the detailed method for obtaining the compressed likelihood).
    \item For \textbf{SN} data, we use the SNe Ia samples of 1701 light curves of 1550 distinct Type Ia supernovae (SNe Ia) in the redshift range $z \in [0.001, 2.26]$ from the latest Pantheon+ compilation \cite{Scolnic:2021amr, Brout:2022vxf}. This sample also includes SH0ES Cepheid host distance anchors (Riess et al, denoted as R21 \cite{Riess:2021jrx}). Thus, for $H_0$, when combining it with Pantheon+, we use the PantheonPlusSH0ES likelihood (refer Eq. 15 of \cite{Brout:2022vxf}), where the Cepheid calibrated host-galaxy distance provided by SH0ES facilitates constraints on $H_0$.
    \item The \textbf{cosmic chronometers} (CC) approach allows us to obtain observational values of the Hubble function at different redshifts $z \leq$ 2 directly. Since these measurements are independent of any cosmological model and Cepheid distance scale, they can be used to place better constraints on it. In the present analysis, we measure $H(z)$ using the CC covariance matrix \cite{M_Moresco_2012, Moresco:2015cya, Moresco:2016mzx}.
    \item For \textbf{Baryon Acoustic Oscillations} (BAO) data, we use data points from the following observations: \\
    Isotropic BAO measurements by 6dF galaxy survey ($z$ $=$ 0.106) \cite{Beutler_2011}, by SDSS DR7-MGS survey ($z$ $=$ 0.15) \cite{Ross:2014qpa} and measurements by SDSS DR14-eBOSS  quasar samples ($z$ $=$ 1.52) \cite{Ata:2017dya}. Measurement of BAO using Ly$\alpha$ samples jointly with quasar samples from the SDSS DR12 study \cite{duMasdesBourboux:2017mrl}. We also consider anisotropic BAO measurements by BOSS DR12 galaxy sample at redshifts 0.38, 0.51, and 0.61 \cite{BOSS:2016wmc}. 
    \item The H$_2$O \textbf{Megamaser} (hereafter, ``MASERS'') technique under the Megamaser Cosmology Project (MCP) leads to the direct measurement of the $H_0$ by measuring angular-diameter distances to galaxies UGC 3789, NGC 5765b, and NGC 4258 in the Hubble flow redshifts $z = 0.0116, 0.0340,$ and $0.0277$ respectively. \cite{Reid_2009, Braatz_2010, Reid_2013, Kuo_2013, Gao_2016}. The technique is based only on geometry, independent of standard candles and the extra-galactic distance ladder, and may provide an accurate determination of H$_0$.
    \item In addition to geometric probes, we use the \textbf{growth rate ($f\sigma_8(z)$)} data provided by various galaxy surveys as collected in \cite{Basilakos:2013nfa} to constrain cosmological parameters.
\end{itemize}
Now, to constrain the free and derived parameters of this coupled cosmological scenario, we use the Markov Chain Monte Carlo (MCMC) technique, emcee: the MCMC Hammer \cite{Foreman-Mackey_2013}. We work with the following parameter space $\mathsf{P} \equiv \Bigl\{ \Omega_{\phi_i}*10^{-9}, r_{drag}, \lambda_i, m_i, \Omega_{r_i}, h, \sigma_8 \Bigr\}$ during the statistical analyses and the priors employed on these cosmological parameters are enlisted in Table \ref{table:flat priors}. The physical limits are intact where $\Omega_{\phi}$ > 0,  $\Omega_{dm}$ > 0 and $\Omega_{r}$ > 0.
 During our further analysis, the Hubble constant is assumed to be $H_0 = 100 h$ km s$^{-1}$ Mpc$^{-1}$; hence, we define the dimensionless parameter $h$. We have fixed
the baryon density parameter, $\Omega_{b_0}$ to be 0.045 according
to CMB constraints from \textit{Planck} (2018) \cite{Planck:2018vyg}, which is also
in agreement with Big Bang Nucleosynthesis (BBN) constraints and $n_s$ to be 0.96. 
One should note that we have chosen positive priors on the interaction parameter. However, the results do not change if we choose negative or bigger priors alternatively.


\section{Results and Discussion}\label{results and discussion}
\begin{table}[!ht]
\small
\centering
\addtolength{\tabcolsep}{8pt} 
\renewcommand{\arraystretch}{1.2} 
\begin{tabular}{|c | c|}
 \toprule
 \textbf{\textit{Parameters}}  & \textbf{\textit{Flat prior interval}} \\  
 \midrule
 $\Omega_{\phi_i}*10^{-9}$ & [0.05, 4.0]  \\  
 $r_{\text{drag}}$ & [130, 180] Mpc \\
 $\lambda_i$ & [$10^{-4}$, 3.0]  \\
 $m_i$ & [$10^{-6}$, $10^{-2}$] \\  
 $\Omega_{\text{r}_i}$ & [0.09, 0.17]  \\  
 $h$ & [0.5, 0.9]  \\
 $\sigma_{8}$ & [0.6, 1.0]  \\  
 $M$ & [-21.0, -18.0] \\
\bottomrule
\end{tabular}
\caption{Flat priors on various parameters of the coupled model.}
\label{table:flat priors}
\end{table}



\begin{table}[!ht]
\small
\centering
\addtolength{\tabcolsep}{-5.0pt} 
\renewcommand{\arraystretch}{2}
\begin{tabular}{ c c c c c c c c c} 
\toprule [1.5pt]
  \textbf{\textit{Parameters}}  &  $\Omega_{\phi_i}*10^{-9}$   & $r_{drag}$(Mpc) & $\lambda_i$  & $m_i$ & $\Omega_{r_i}$ & $\Omega_{dm}$ & $h$ & $\sigma_{8}$\\
\bottomrule [1.5pt]

 \multicolumn{9}{c}{ CMB+CC+BAO+MASERS} \\
 \hline
 \textbf{\textit{Coupled}} & 1.125$^{+0.275}_{-0.264}$ & 147.73$^{+5.83}_{-5.86}$ & 0.99$^{+0.46}_{-0.82}$ & 0.0010$^{+0.00038}_{-0.00096}$ & 0.151$^{+0.0102}_{-0.0105}$  & 0.317$^{+0.0121}_{-0.0126}$   & 0.671$\pm$0.028 &  Unconstrained \vspace{0.15cm}  \\ 
\textbf{\textit{Un-coupled}} & 1.448$\pm$0.053 & 150.28$^{+5.61}_{-5.79}$ & 0.64$^{+0.447}_{-0.436}$ & - & 0.153$^{+0.0096}_{-0.0099}$ & 0.311$^{+0.0084}_{-0.0085}$   & 0.665$\pm$0.026 & Unconstrained \vspace{0.2cm} \\ 
 \bottomrule [1.3pt]
  \multicolumn{9}{c}{ CMB+PantheonPlus+CC+BAO+MASERS} \\ 
 \hline  
 \textbf{\textit{Coupled}} &  1.073$^{+0.285}_{-0.280}$ & 148.53$^{+6.21}_{-6.27}$ & 0.83$^{+0.46}_{-0.32}$ & 0.0011$^{+0.00040}_{-0.00110}$ & 0.151$^{+0.0101}_{-0.0103}$  & 0.314$\pm$0.0074   & 0.671$^{+0.028}_{-0.027}$ &  Unconstrained \vspace{0.15cm} \\ 
\textbf{\textit{Un-coupled}} & 1.441$\pm$0.054 & 150.30$^{+5.69}_{-5.75}$ & 0.70$^{+0.414}_{-0.384}$ & - & 0.153$^{+0.0095}_{-0.0098}$ & 0.313$\pm$0.0072   & 0.665$^{+0.026}_{-0.025}$ & Unconstrained  \vspace{0.2cm}\\ 
 \bottomrule [1.3pt]
 
 \multicolumn{9}{c}{ CMB+PantheonPlus+CC+BAO+MASERS+$f\sigma_8(z)$} \\
 \hline
 \textbf{\textit{Coupled}} &  1.041$^{+0.295}_{-0.290}$ & 148.31$^{+6.01}_{-5.95}$ & 0.80$^{+0.49}_{-0.34}$ & 0.0012$^{+0.00047}_{-0.00110}$ & 0.150$^{+0.0101}_{-0.0101}$  & 0.314$^{+0.0074}_{-0.0075}$  & 0.672$\pm$0.027 & 0.740$^{+0.022}_{-0.023}$ \\ 
\textbf{\textit{Un-coupled}} & 1.441$\pm$0.053 & 150.30$^{+5.64}_{-5.79}$ & 0.71$^{+0.411}_{-0.373}$ & - & 0.153$^{+0.0096}_{-0.0097}$ & 0.313$^{+0.0073}_{-0.0072}$   & 0.665$^{+0.026}_{-0.025}$ & 0.740$\pm$0.022   \\ 
 \bottomrule [1.3pt]
 
 \multicolumn{9}{c}{ CMB+PantheonPlus+CC+BAO+MASERS+$f\sigma_8(z)$+$H_0$} \\
 \hline
 \textbf{\textit{Coupled}} &  1.086$^{+0.288}_{-0.284}$ & 139.03$^{+3.16}_{-3.15}$ & 0.79$\pm$0.43 & 0.0011$^{+0.00046}_{-0.00110}$ & 0.134$^{+0.0057}_{-0.0056}$  & 0.313$^{+0.0085}_{-0.0086}$   & 0.717$\pm$0.015 & 0.741$\pm$0.022 \\ 
\textbf{\textit{Un-coupled}} & 1.476$\pm$0.052 & 139.85$^{+3.08}_{-3.09}$ & 0.52$^{+0.374}_{-0.380}$ & - & 0.135$^{+0.0057}_{-0.0057}$ & 0.310$^{+0.0075}_{-0.0076}$   & 0.716$\pm$0.015 & 0.740$\pm$0.022   \\
 \bottomrule [1.3pt]
\end{tabular}
{\centering\caption{\label{mean with error} Mean values with 68\% confidence level (CL) errors on cosmological and free parameters within the interacting and non-interacting paradigm from various data combinations.}}
\end{table}
\begin{figure}[ht!]
    \centering
    \includegraphics[width=1.0\textwidth]{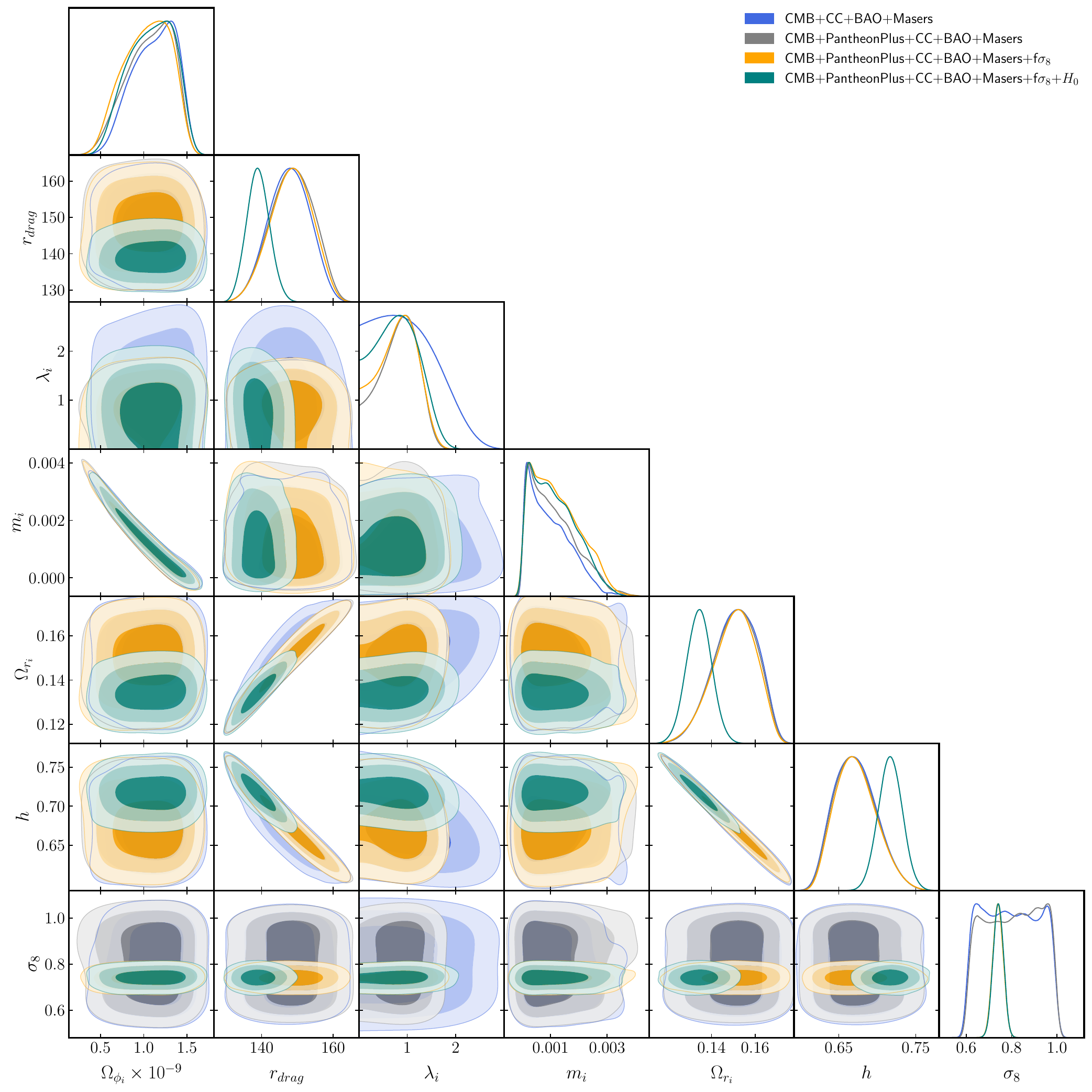}
    \caption{68$\%$ and 95$\%$ CL contour plots and corresponding one-dimensional marginalized posterior distribution for all cosmological parameters obtained from the MCMC analysis within the present \textbf{coupling} scenario utilizing several combinations of data sets.}
    \label{fig:merging_coupled}
\end{figure}
\begin{figure}[ht!]
    \centering
    \includegraphics[width=1.0\textwidth]{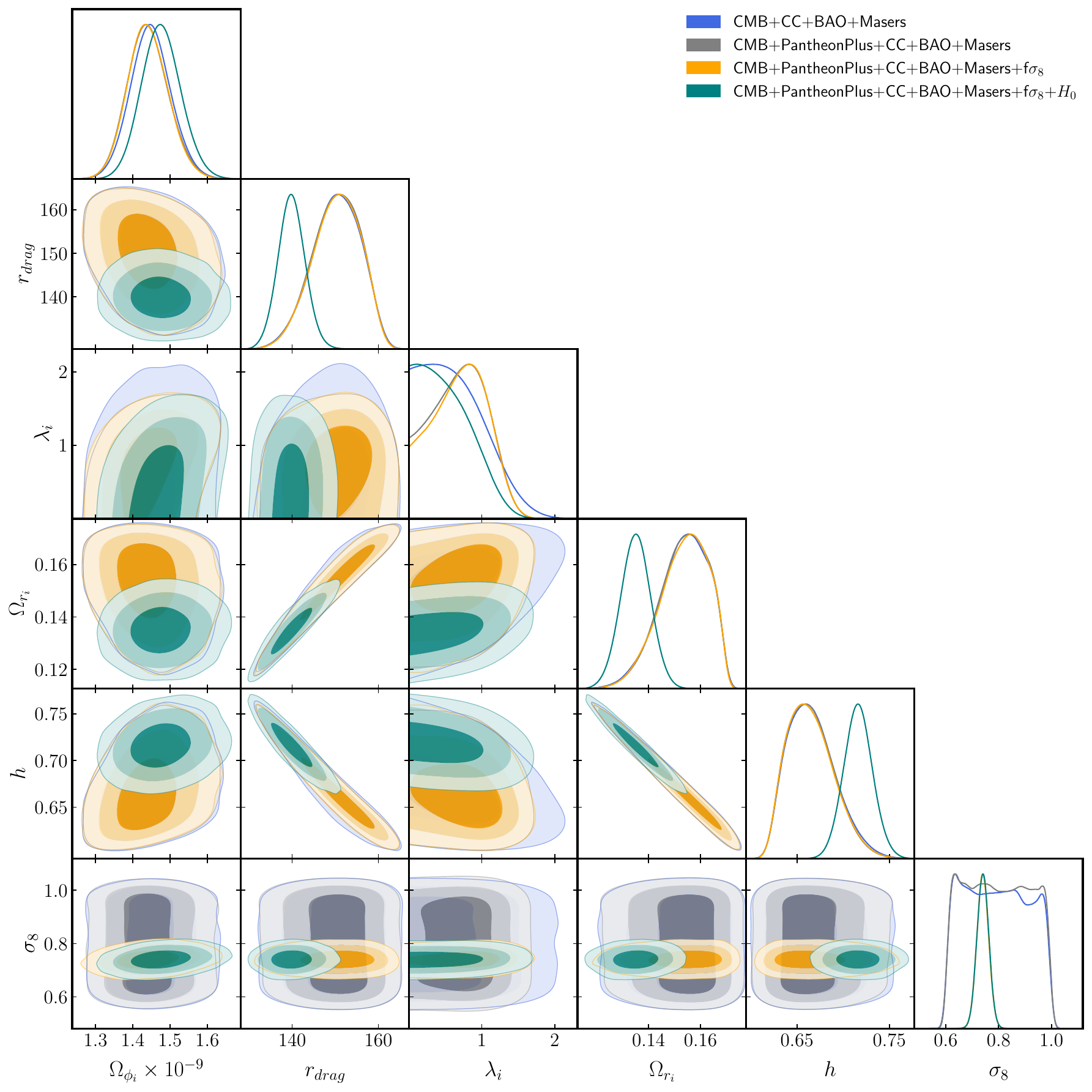}
    \caption{68$\%$ and 95$\%$ CL contour plots and corresponding one-dimensional marginalized posterior distribution for all cosmological parameters obtained from the MCMC analysis for \textbf{non-coupling} scenario utilizing several combinations of data sets.}
    \label{fig:merging_uncoupled}
\end{figure}
\begin{table}[ht!]
\small
\centering
\renewcommand{\arraystretch}{2}
\addtolength{\tabcolsep}{-2.3pt} 
\begin{tabular}{c c | c c c c c} 
\toprule [1.2pt]
&{\textbf{\textit{Parameters}}}  &  $\Omega_{dm}$   & $r_{drag}$(Mpc) & $h$ & $\Omega_{r_0}*10^{-5}$ & $\sigma_{8}$\\ \cline{2-7}
\multicolumn{1}{c}{\multirow{5}{*}{\begin{sideways}\textbf{\textit{ Data combinations}} \end{sideways}}}
&\makecell{CMB+CC+BAO+MASERS} & 0.308$^{+0.0078}_{-0.0076}$ & 149.34$^{+6.25}_{-6.24}$ & 0.674$\pm$0.028 & 5.41$^{+0.398}_{-0.400}$ & Unconstrained \vspace{0.10cm}\\  \cline{2-7} 
&\makecell{CMB+PantheonPlus+CC+BAO+\\MASERS} & 0.312$^{+0.0069}_{-0.0070}$ & 149.16$^{+6.36}_{-6.34}$ & 0.673$\pm$0.028 & 5.45$^{+0.408}_{-0.403}$ & Unconstrained \vspace{0.10cm}\\ \cline{2-7} 
&\makecell{CMB+PantheonPlus+CC+BAO+\\MASERS+$f\sigma_8(z)$} & 0.312$^{+0.0070}_{-0.0071}$ & 149.42$^{+6.49}_{-6.48}$ & 0.672$^{+0.029}_{-0.028}$ & 5.46$^{+0.407}_{-0.415}$ & 0.736$\pm$0.022 \vspace{0.10cm}  \\ \cline{2-7}
&\makecell{CMB+PantheonPlus+CC+BAO+\\MASERS+$f\sigma_8(z)$+$H_0$} & 0.309$\pm$0.0073 & 139.60$^{+3.16}_{-3.12}$ & 0.719$\pm$0.015 & 4.79$^{+0.219}_{-0.218}$ & 0.738$\pm$0.022 \vspace{0.10cm} \\
 \bottomrule[1.2pt]
\end{tabular}
{\centering\caption{\label{mean with error LCDM} Mean values with 68\% confidence level (CL) errors on cosmological and free parameters within the $\Lambda$CDM paradigm from various data combinations.}}
\end{table}
We start by discussing the results obtained within the interacting canonical context. In table \ref{mean with error}, we present the mean values and the 1$\sigma$ ranges of different cosmological parameters with both coupled and uncoupled scenarios. 
\par
To see the effect of the inclusion of each data set in the coupled scenario on the constraint of each parameter, we refer to Fig.\ref{fig:merging_coupled}, where we show 1-dimensional marginalized posterior distributions and the corresponding 2D contour plots at 68$\%$ and 95$\%$ confidence level (CL) for all independent parameters. We gradually added PantheonPlus, $f\sigma_8$, and $H_0$ to the data set combination BAO + Masers + CMB + CC (further renamed as BASE combination) and can see more constrained parameters by adding more datasets. We also see a similar effect in the non-coupling case, as shown in Fig. \ref{fig:merging_uncoupled}.
Since constraints on parameters show not much difference in the case of coupled or un-coupled scenarios, we, from now onward, focus on further studying the coupled case.
\subsection{\textit{The \texorpdfstring{$H_0$}{a} and \texorpdfstring{$\sigma_8$}{T} Plane}}
We see that 1$\sigma$ confidence level contours are relatively larger for CMB + CC + BAO + Masers (BASE) data than other data combinations. Whereas the addition of PantheonPlus sample significantly reduces the constraint on parameter $\Omega_{dm}$, with the difference of $\sim$ 0.1$\sigma$ to the BASE combination. This alteration estimates slightly higher uncertainty in $r_{drag}$, differing by less than $0.1\sigma$ relative to the BASE data. We also observe the change in the $\lambda_i$ parameter with less than $0.3\sigma$ significance compared to the BASE combination, refer Table \ref{mean with error}. Furthermore, the inclusion of $f\sigma_8$ and $H_0$ data to the previous data combination strongly reduces the parameter space on $\sigma_8$ and $h$, respectively, as seen in Fig. \ref{fig:merging_coupled}. Except parameter $h$ and parameter $\sigma_8$, we also find the notable impact on other parameters such as $r_{drag}$ and $\lambda_i$ when we add SH0ES data or when we include $f\sigma_8$ data in the analysis. 

In this study, we found that CMB + CC + BAO + Masers (BASE) data combination shows $H_0$ = 67.1$\pm$2.8 km/s/Mpc at 68$\%$ CL which is compatible well within 0.3$\sigma$ range with $H_0$ measurement from the joint result using TT,TE,EE + lowE + lensing in \textit{Planck} (2018) under $\Lambda$CDM ($H_0$ = 67.37 $\pm$ 0.54 km/s/Mpc) \cite{Planck:2018vyg}. But, the $H_0$ measurement from the same data combination shows 2$\sigma$ discrepancy with the prediction from SH0ES measurement in R21 ($H_0$ = 73.04 $\pm$ 1.01 km/s/Mpc) \cite{Riess:2021jrx}. The consequent addition of PantheonPlus data and the analysis combining $f\sigma_8$ data produces essentially the same result for the Hubble constant as in the BASE combination. However, with the addition of $H_0$ data from SH0ES, we reached a determination of $H_0$ = 71.7$\pm$1.54 km/s/Mpc, 2.5$\sigma$ increase in the difference from \textit{Planck} (2018)+$\Lambda$CDM of 67.37 $\pm$ 0.54 km/s/Mpc. The combined result of $H_0$, 71.7$\pm$1.54 km/s/Mpc, is consistent within < 1$\sigma$ (mildly greater than 0.5$\sigma$) range with the measurement of the Hubble constant, $H_0$ = 73.04 $\pm$ 1.01 km/s/Mpc derived from the baseline fit with measurements from the latest SH0ES analysis in R21 \cite{Riess:2021jrx}.

We now present our multi-probe constraints on the structure growth parameter $S_8$, using $S_8$ = $\sigma_8$ $\sqrt{\Omega_{dm_0}/0.3}$. As evident from Table \ref{mean with error}, the $\sigma_8$ measurements from growth rate ($f\sigma_8$) data, along with CMB, SNIa, CC, BAO, MASERS, and $H_0$ data, are well constrained as compared to datasets that do not include $f\sigma_8$, i.e., ``CMB + CC + BAO + MASERS'' and ``CMB + PantheonPlus + CC + BAO + MASERS'' combinations. This discrepancy in $\sigma_8$ significantly impacts the reliability of the constraints on parameter $S_8$ for these data combinations (without $f\sigma_8$) because obtained $S_8$ solution arises from the parameter being unconstrained. Due to this,  we refrain to present the constraints placed on $S_8$ for these combinations.
\par
Advancing with $S_8$ = $\sigma_8$ $\sqrt{\Omega_{dm_0}/0.3}$, the addition of $f\sigma_8$ measurement yields systematically low value of $S_8$ and resulted in $S_8$ = 0.758$\pm$0.023 raising the difference with \textit{Planck} (2018)+$\Lambda$CDM ($S_8$ = 0.830 $\pm$ 0.013) to precisely at 2$\sigma$. This prediction of $S_8$ is in excellent agreement with the KiDS-1000-BOSS results \big($S_8$ = 0.766$^{+0.020}_{-0.014}$ \big) \cite{Heymans:2020gsg}, differing by 0.3$\sigma$. Whereas the estimation is in good agreement with DES-Y3 results \big($S_8$ = 0.776$\pm$0.017\big) \cite{PhysRevD.107.023531} with the difference of 0.5$\sigma$. The results from CMB + CC + BAO + Masers + PantheonPlus + $f\sigma_8(z)$ are more robust in tightening constraints on the $\sigma_8$ or the $S_8$ parameter and justify the reduced uncertainty and the parameter space compared to the results from without growth rate data combination; see Table \ref{mean with error} and Fig.\ref{fig:merging_coupled}. Finally, the combined data constraint yields $S_8$ = 0.757$\pm$0.023, a difference of 2.5$\sigma$ from the results of \textit{Planck} (2018)+$\Lambda$CDM \cite{Planck:2018vyg}. However, this analysis of combined data produces essentially the same results as in the previous analysis maintaining a good agreement of 0.3$\sigma$ and 0.5$\sigma$ with the measurements obtained from KiDS-1000-BOSS survey \cite{Heymans:2020gsg} and DES-Y3 survey \cite{PhysRevD.107.023531}, respectively. 

To assess the robustness of our findings and to check for the ambiguity concerning the prior dependency in the obtained results we present the results with different priors on the cosmological parameters in Table \ref{mean with error_sigma8_0.5-1} and in Table \ref{appendix_meanwitherror_differentpriors}, in the Appendix. We found that the values of the key cosmological parameters (both mean and error bars) are independent of priors only when the $f\sigma_8$ is in consideration along with other data sets of CMB, SNIa, CC, BAO, MASERS, and $H_0$. This is explained in detail in the Appendix.

\subsection{\textit{The \texorpdfstring{$h$} \texorpdfstring{-}\texorpdfstring{$r_{drag}$}{P} Plane}}
In addition to a discrepancy in
the Hubble constant, there is a discrepancy in the co-moving sound horizon at the end of the baryon drag epoch, $r_{drag}$ as well. The two parameters $H_0$ and $r_{drag}$ are strictly related when we consider BAO observations. In actuality, a combination of expansion history probes such as BAO and PantheonPlus data can provide a model-independent estimate of the low-redshift standard ruler, constraining the product of $h$ (with $H_0$= $h$ × 100 km/s/Mpc) and the sound horizon $r_{drag}$ directly. This implies that, to have a higher $H_0$ value in agreement with SH0ES, we need $r_{drag}$ $\sim$ 137 Mpc, while to agree with Planck, assuming $\Lambda$CDM, we need $r_{drag}$ $\sim$ 147 Mpc. For this reason, the solutions that increase the expansion rate and at the same time decrease $r_{drag}$ are most promising. In our analysis, this feature is completely in agreement with the \textit{Planck} (2018) estimates due to the strong compatibility of estimated $H_0$ with the one from \textit{Planck} (2018) + $\Lambda$CDM model. However, the addition of SH0ES result gives $H_0$ = 71.7$\pm$1.54 km/s/Mpc, that is closer to the R21 value of $H_0$, also leads to an empirical determination of $r_{drag}$ near 139 Mpc. This modest discrepancy in the sound horizon value $r_{drag}$ from the one in \cite{Bernal:2016gxb} agrees with the slight discrepancy in the $H_0$ measurements from SH0ES. This correlation can also be seen from table \ref{mean with error}. Evidently, this cosmological solution is promising and consistent with the fact that the relation $h$ - $r_{drag}$ is constant by the BAO measurements.
\subsection{\textit{The \texorpdfstring{$S_8$} \texorpdfstring{-} \texorpdfstring{$\Omega_{dm}$}{P} Plane}}
\begin{figure}
    \centering
    \begin{minipage}[t]{0.48\textwidth}
        \centering
        \includegraphics[width=1.03\textwidth]{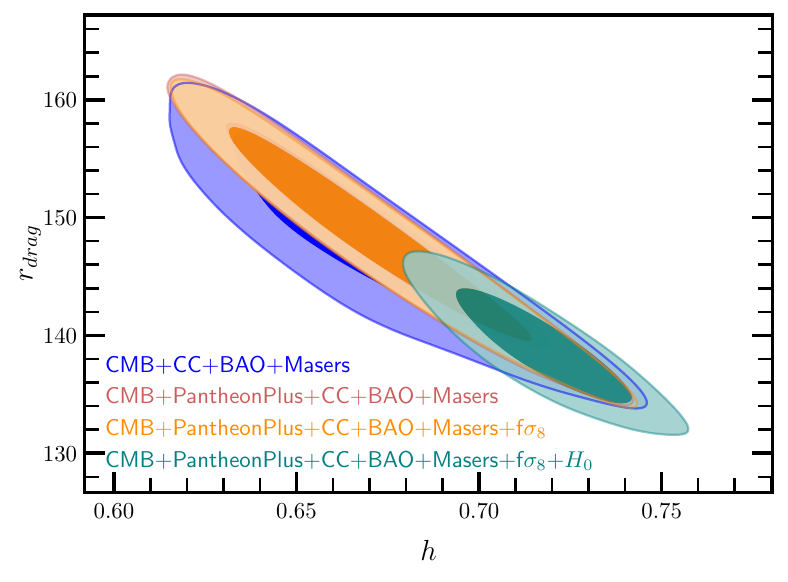} 
        \caption{Constraints in the $h$-$r_{drag}$ plane from different combination of CMB, CC, BAO, Masers, Pantheon, growth and $H_0$ data (68\% and 95\% contours) in case of interacting model. }
    \label{fig:h0-rdrag_plot}
    \end{minipage}\hfill
    \begin{minipage}[t]{0.48\textwidth}
        \centering
        \includegraphics[width=1.03\textwidth]{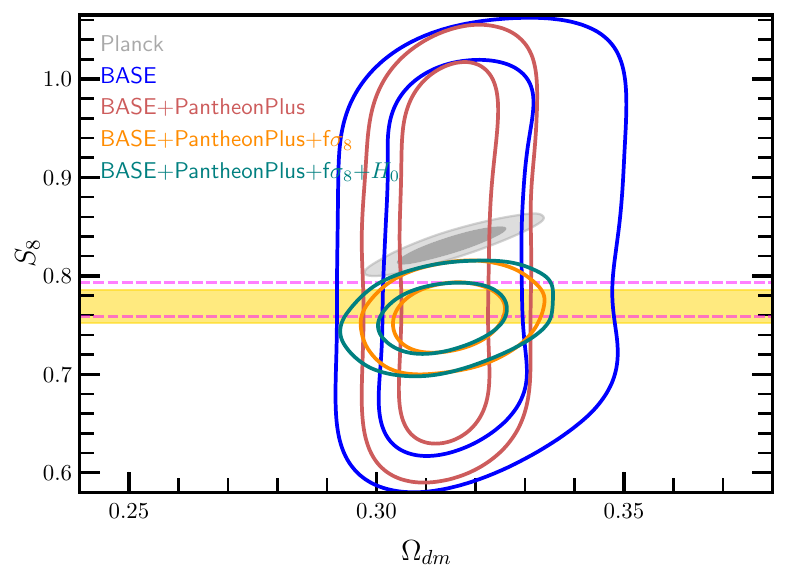} 
        \caption{Marginalised constraints for the joint distributions of the fluctuation amplitude parameter $S_8$ and dark matter density parameter $\Omega_{dm}$. The 68\% and 95\% credible regions are shown for different sets of CMB, SNIa, CC, BAO, MASERS, $H_0$, and growth data for an interacting model. The golden band reflects the 68$\%$ credible interval for $S_8$ from the KiDS-1000-BOSS joint analysis whereas the region between pink dashed lines represent  DES-Y3 results. For comparison, we also incorporated the \textit{Planck} (2018) results (in grey contour). The BASE results are for ``CMB + CC + BAO + Masers'' dataset.}
  \label{fig:combined_sigma8-odm}
    \end{minipage}
\end{figure}

The model that allows larger $H_0$ values tend to introduce other tensions, such as higher $S_8$ values. To obtain simultaneously higher values of $H_0$, lower values of $S_8$, and consistent values of $\Omega_{dm}$ is also necessary to define the correct evolution. Since adding the BAO and SNe measurements to the \textit{Planck} data strengthens the constraints towards \textit{Planck} values, the correlation between their combined results is relatively strong. Referring to the Table \ref{mean with error}, $f{\sigma_8}$ addition improves the $S_8$ consistency with KiDS-1000-BOSS \cite{Heymans:2020gsg} and DES-Y3 \cite{PhysRevD.107.023531} results. This is also illustrated in Fig. \ref{fig:combined_sigma8-odm}. We, particularly, show that only dataset that includes $f{\sigma_8}$, constrain $S_8$ or $\sigma_8$ better while other datasets (without $f{\sigma_8}$) fails to put constrain on it. From this joint analysis, we found a lower and a well-constrained estimation for $S_8$, leading to a lower and a constrained value for $\Omega_{dm}$ (Table \ref{mean with error} and Fig. \ref{fig:combined_sigma8-odm}). Moreover, using the full combined likelihood for the data considered in this work we found the similar results with further increment in the $H_0$ estimated value
while $S_8$ retaining the consistency with KiDS-1000-BOSS and DES-Y3 survey results.

\subsection{\textit{The Co-moving Hubble Parameter}}
\begin{figure}
    \centering
    \begin{minipage}[t]{0.48\textwidth}
        \centering
        \includegraphics[width=1.06\textwidth]{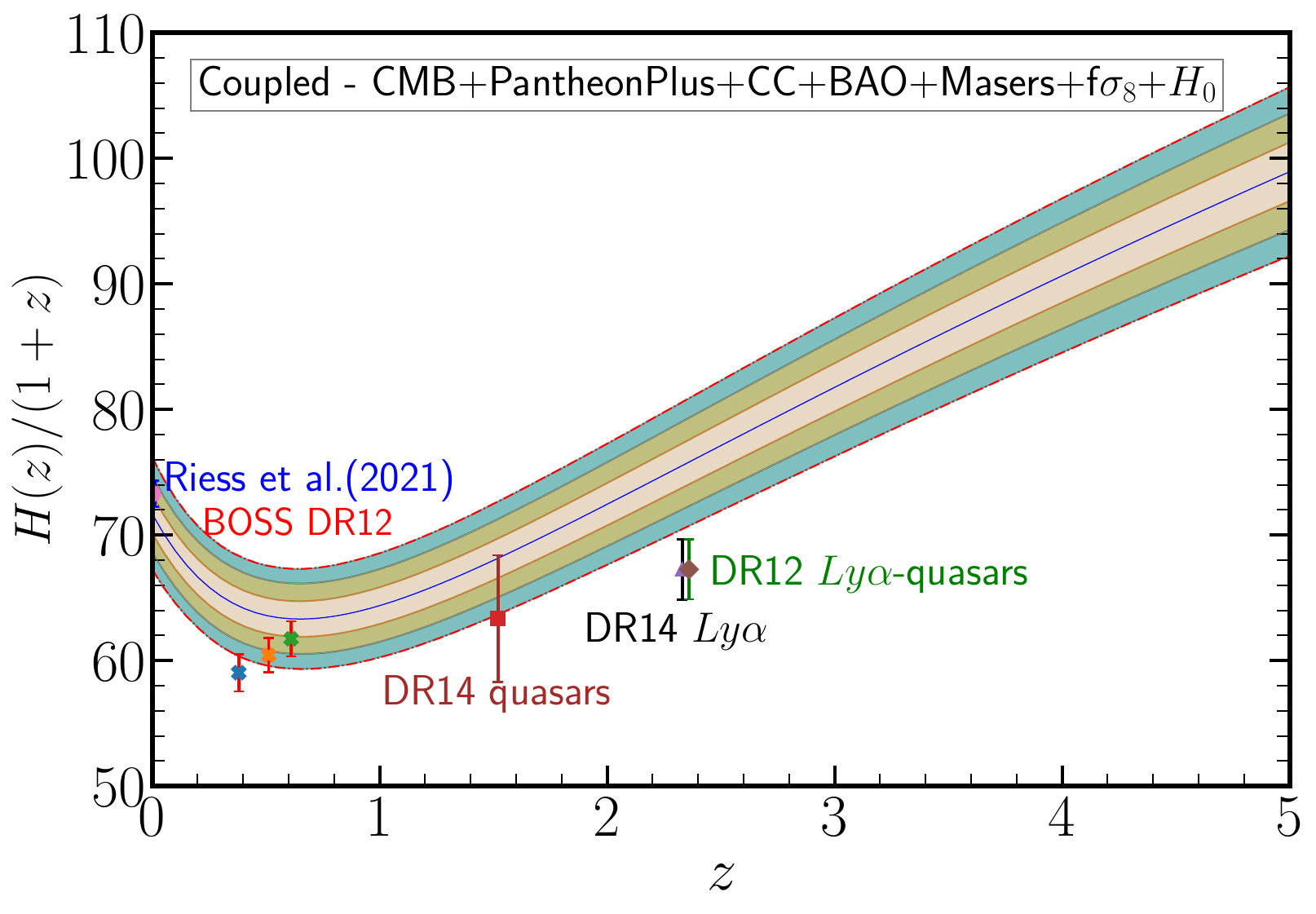} 
        \caption{Co-moving Hubble parameter as a function of redshift, clearly showing the onset of acceleration just before the redshift around $z$ = 0.6, reconstructed using a combination of CMB, SNIa, CC, BAO, Masers, $H_0$ and growth data for interacting model. The solid blue line shows the best-fit reconstructed
results, while the outer regions show results within 1$\sigma$, 2$\sigma$, 3$\sigma$ CL separated by orange, pink and red-dashed lines.}
    \label{fig:Hz_1+z}
    \end{minipage}\hfill
    \begin{minipage}[t]{0.48\textwidth}
        \centering
        \includegraphics[width=0.97\textwidth]{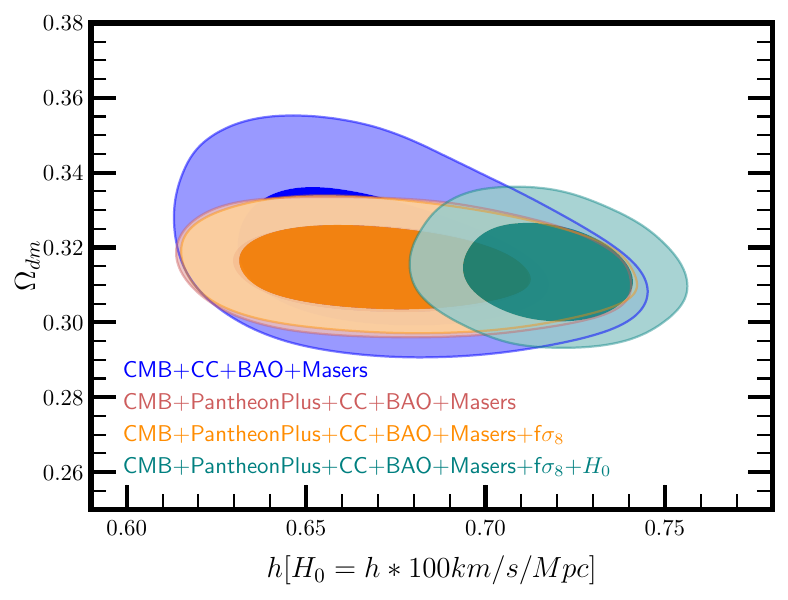} 
        \caption{The CMB + SNIa + CC + BAO + Masers + $f{\sigma_8}$ + $H_0$ data constraints on $h$ and $\Omega_{dm}$ in the coupled model, compared to the results from without $H_0$ data. Adding SH0ES $H_0$ constrains the Hubble parameter, comparable to that R21 estimation. Darker and lighter contours show 68\% and 95\% of the probability, respectively.}
        \label{fig:h0-omega_dm}
    \end{minipage}
\end{figure}

The recent BAO measurements along the line of sight and transverse directions lead to joint constraints on $H(z) r_{drag}$.
Since \textit{Planck} + $\Lambda$CDM constrains $r_{drag}$ to a precision of 0.2 \%, the BAO measurements can be accurately converted into absolute measurements of $H(z)$.
These constraints on $H(z)$ from the BOSS analyses are plotted in Fig. \ref{fig:Hz_1+z}. The error bars are constraints from BOSS DR12 galaxy sample \cite{BOSS:2016wmc}, eBOSS DR14 quasar sample \cite{Ata:2017dya}, the correlations of
Ly$\alpha$ absorption in eBOSS DR14  at $z=2.34$ \cite{deSainteAgathe:2019voe} and from the Ly$\alpha$ auto-correlation and cross-correlation with quasars from SDSS data release DR12 \cite{duMasdesBourboux:2017mrl}. The error bar at $z = 0$ shows the inferred distance-ladder Hubble measurement from R21 \cite{Riess:2021jrx}.
\par
The illustration in Fig. \ref{fig:Hz_1+z} shows clearly how well the dark energy interaction model fits the BAO measurements of the Hubble parameter except for the DR12 $Ly$-$\alpha$ data point. It is also consistent with the R21 measurements of Hubble at $z=0$ for all the data sets combined in this work within 1$\sigma$ error bars.  This is also illustrated in Fig. \ref{fig:h0-omega_dm}, which shows the combined constraints on $h$ and $\Omega_{dm}$ from different data combinations. Without adding $H_0$ to the combination of CMB, BAO, PantheonPlus, $H(z)$, Masers, and $f{\sigma_8}$, the tension with SH0ES measurement of $H_0$ still prevails as seen in table \ref{mean with error}, though the same results show lower mean value and well improved constraints on $\Omega_{dm}$ as compared to the BASE data.  However, the addition of SH0ES data (green contour) constrain $\Omega_{dm}$ towards slightly lower mean value and shifts $H_0$ closer to the R21 measurement.\\
\subsection{\textit{Other Cosmological Parameters}}
\begin{figure}
    \centering
    \begin{minipage}[t]{0.48\textwidth}
        \centering
        \includegraphics[width=0.98\textwidth]{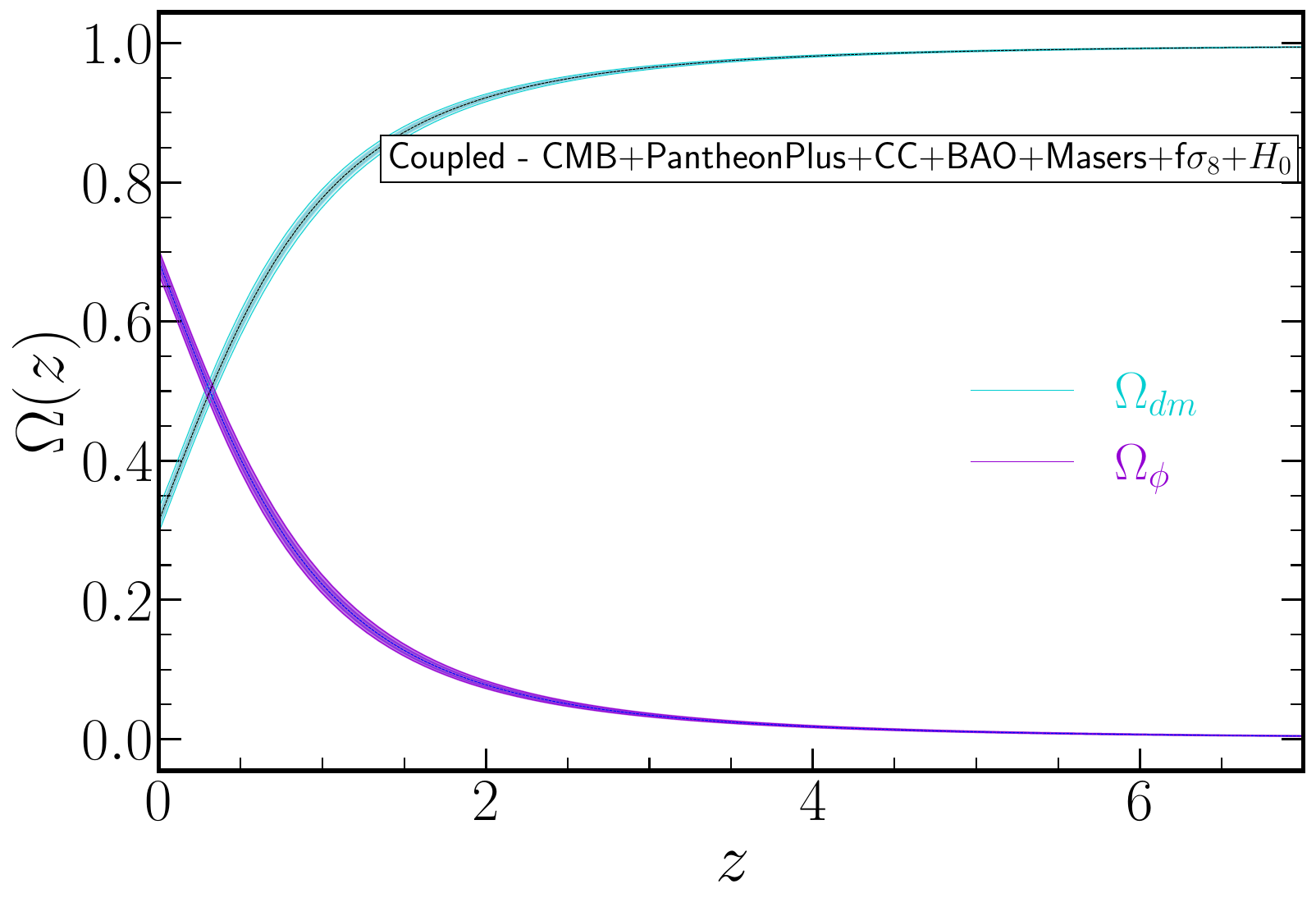} 
        \caption{The 68\% and 95\% CL along with the mean value for relative energy density parameters $\Omega_{dm}$ (cyan) and $\Omega_{\phi}$ (purple) for the coupled model with respect to redshift $(z)$. The evolution is plotted for all the data combined together, considered in this work. The black-dashed and blue-dashed lines corresponds to the best-fit values of DM and DE density parameters.}
    \label{fig:density_evolution}
    \end{minipage}\hfill
    \begin{minipage}[t]{0.48\textwidth}
        \centering
        \includegraphics[width=1.02\textwidth]{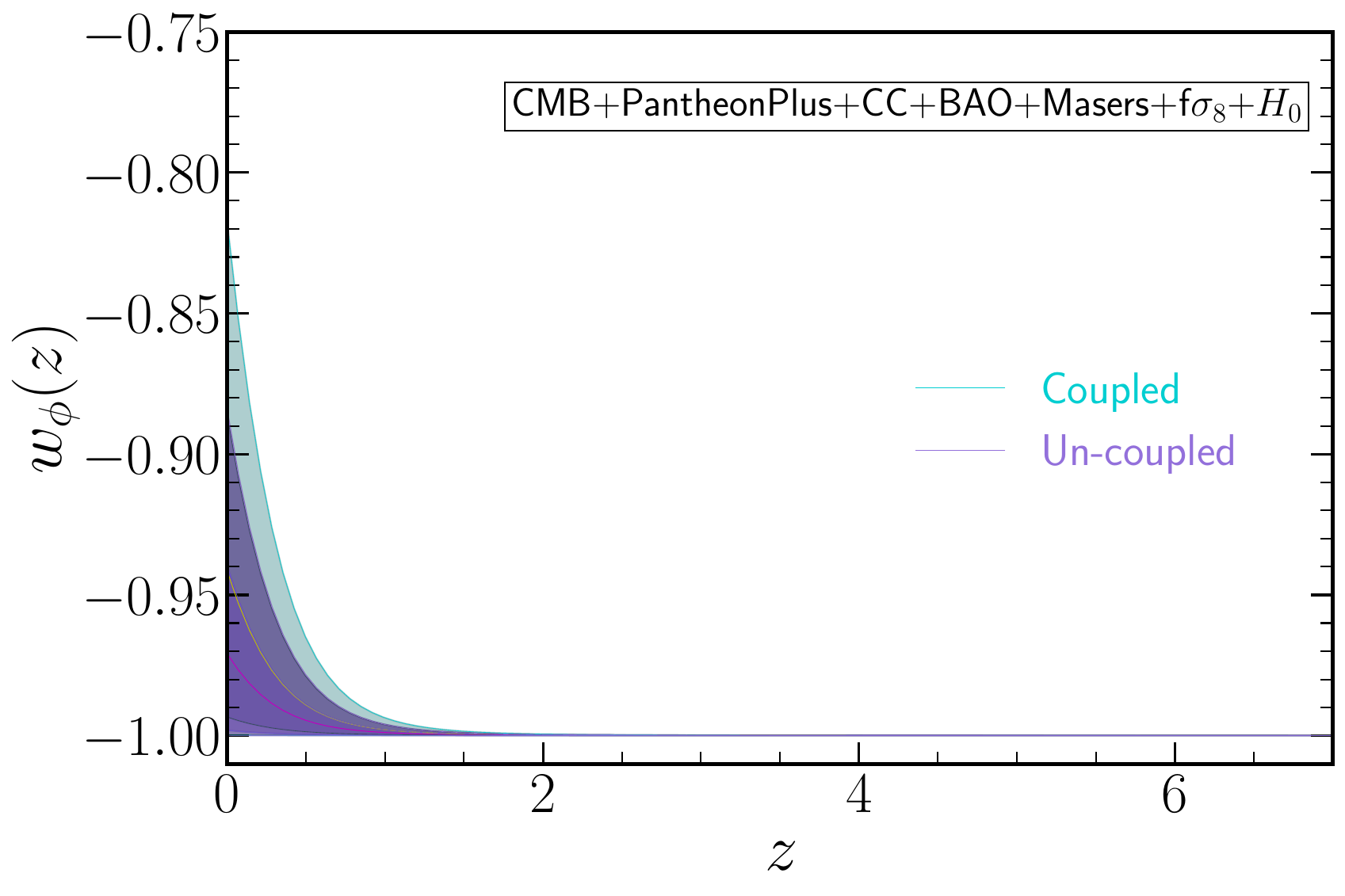} 
        \caption{Evolution of dark energy equation of state parameter with respect to redshift $(z)$ for coupled and un-coupled scenario. The solid yellow line shows the best fit result while the outer blue and cyan regions show the 1$\sigma$ and 2$\sigma$ CL for coupled case. The solid red line shows the best fit result while the outer purple regions show the 1$\sigma$ and 2$\sigma$ CL for un-coupled case.}
        \label{fig:wphi_overlapping}
    \end{minipage}
\end{figure}

\begin{SCfigure}
    \includegraphics[width=0.55\textwidth]{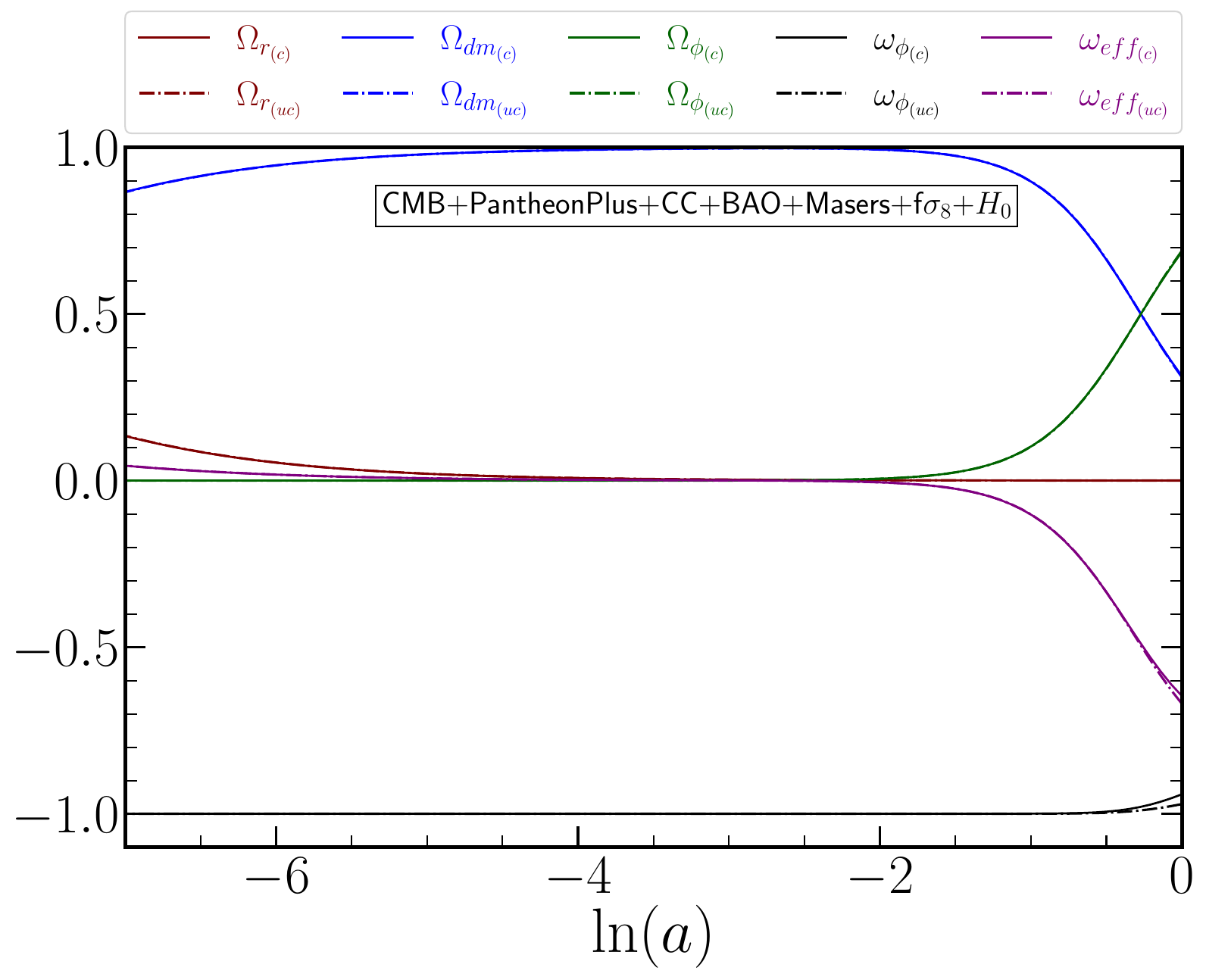}
  \caption{The best-fit evolution of key cosmological parameters for the coupled and uncoupled models is shown for the combined set of  CMB, SNIa, CC, BAO, Masers, $H_0$, and growth data. The solid lines and dashed-dot lines show the evolution in the coupled case and un-coupled case, respectively.}
  \label{fig:total_all_parameter}
\end{SCfigure}

We further analyse the model for completeness' sake and clarity. We have plotted the energy density evolution parameter for dark energy and dark matter, along with 68\% and 95\% CL for the coupled case using all the data presented in this work in Fig. \ref{fig:density_evolution}. Fig. \ref{fig:wphi_overlapping} shows the overlapping confidence contours for the equation of state parameter as a function of redshift from coupled and uncoupled models using all the data presented in this work. The solid yellow and red lines indicate the best-fit values for the DE EoS parameter for coupled and uncoupled models, respectively. However, Fig. \ref{fig:total_all_parameter} shows the density parameters and equation of state parameters evolution from their best-fit values as a function of the logarithmic scale factor. These analyses exactly show the expected behaviour of the observed universe at present, which is dark energy-dominated accelerated expansion. Moreover, while fitting with the data, we found that the model favors small values of the coupling parameter $``$Q'', making the parameter evolution indistinguishable compared to the one from the uncoupled case. However, these predictions are model-dependent and may vary for other coupling models.

\subsection{\textit{Comparison with \texorpdfstring{$\Lambda$CDM}{r} results}} 
In Table 3, we present the constraints that we got from $\chi^2$ minimization fitting assuming the standard $\Lambda$CDM model. We also show how the mean values change within the same model in addition to different datasets. We observe the notable impact on the addition of PantheonPlus sample to the BASE combination that significantly changes the parameter, $\Omega_{dm}$ values (both mean and error bars). These changes narrow down the $\Omega_{dm}$ parameter space. Growth rate ($f\sigma_8(z)$) data addition reflects more precised constraint on $\sigma_8$ or $S_8$ parameter compared to ``CMB + PantheonPlus + CC + BAO + Masers'' data set, keeping other parameters unaltered. Further inclusion of local $H_0$ value from SH0ES 
shifts the mean value of $h$ towards Riess et al. value \cite{Riess_2022} with smaller error bars, differing by $\sim$ $1.5\sigma$ relative to the $h$ value obtained from without SH0ES data sets. The effect of the shift in $h$ can also be seen in $r_{drag}$ parameter values (both mean and reduced error bars) and in $\Omega_{dm}$ parameter value with lower mean.

\subsection{\textit{Model Selection Statistics}}
In this section, we discuss the comparison of our models (coupled and uncoupled) with the standard $\Lambda$CDM model. The $\Lambda$CDM model corresponds to the dark energy equation of state -1, whereas in our model, the dark energy equation of state is allowed to vary. One way to compare these dark energy models is Akaike Information Criterion (AIC), where AIC is defined as AIC = $\chi^2_{min}$ + 2 $\cdot$ d. Another way is Bayesian Information Criterion (BIC), where BIC is defined as BIC = $\chi^2_{min}$ + d $\cdot$ ln$(N)$. $\chi^2_{min}$ is the chi-squared minimization to measure how well the data fit a model. $`$d' is the number of parameters in the model and `$N$' is the number of observational data points. The lowest chi-squared and lowest AIC or BIC at each point indicate where the parameters and model most closely match the measured data. Comparing both the methods, we can see from table \ref{table:model comparison} that there is always an improvement in $\chi^2_{min}$ value for the coupled model when compared to the un-coupled or $\Lambda$CDM model. For all the data together, this leads to a large reduction in AIC and BIC estimates for the coupled model, and hence, the same is significantly preferred over the $\Lambda$CDM model.
\par
We also considered $\chi^2_{min}$ per number of degrees of freedom approach for the present model, aiming to understand the observational solidity of the model with respect to the standard reference model. We found that $\frac{\chi^2_{min}}{n_{dof}}$ raises an over-fitting issue with the coupled model opposing the analysis from AIC. This trait can be found in Table \ref{table:model comparison}. However, using $\frac{\chi^2_{min}}{n_{dof}}$ method, it is clear that the $\Lambda$CDM model has the goodness of fit (GoF) to the data and hence, is the favoured model. In summary, from the point of view of
the $\frac{\chi^2_{min}}{n_{dof}}$ analysis, $\Lambda$CDM is still a preferred
candidate but from the perspective of AIC and BIC analyses, coupled cosmology is the preferred scenario for the universe’s evolution.

\begin{table}[!htb]
\small
\centering
\renewcommand{\arraystretch}{2}
\begin{tabular}{c c c c c c c c c}
 \hline
 \textbf{\textit{Model}}  & $\chi^2_{min}$ & $\frac{\chi^2_{min}}{n_{dof}}$ &  d  & N & AIC & BIC & $\Delta$ AIC &  $\Delta$ BIC \\  
 \hline\hline
 \textbf{\textit{Coupled}} & 1216.063 & 0.699 & 8 & 1748 & 1232.063  & 1275.792  & 373.365 & 362.433\\
 \textbf{\textit{Un-coupled}} & 1345.614 & 0.773 & 7 & 1748 & 1359.614  &  1397.877 & 245.814 & 240.348\\
 \textbf{\textit{$\Lambda$CDM}}  & 1593.428 & 0.912 & 6 & 1748 & 1605.428  &  1638.225 & 0.0   & 0.0 \\ 
\bottomrule
\end{tabular}
\caption{The $\chi^2_{min}$, AIC and BIC estimation for coupled/un-coupled and $\Lambda$CDM model for the combination of all data set.}
\label{table:model comparison}
\end{table}

\section{Final Remarks}\label{final remarks}
In this work, we study an interacting dark energy-dark matter model having time-varying interaction term $Q = F,_{\phi} \rho_{dm}\dot{\phi}$. This work is interesting because it is theoretically very well motivated from a field theory perspective, not just phenomenological. We can see that because of the time-varying interaction term, the dark energy component may differ from the cosmological constant at very early times ($z= 1100$). Also, a small interaction is not disfavoured by the data at late times. However, we observed that there is no significant difference in the constraining power of the coupled model over the uncoupled one. We found that the parameter $m_i$ which is the coupling parameter peaks at the very lower value in the prior range provided, which makes the coupling significantly low making it indistinguishable from the uncoupled case.
\par
Initially, we tested our model to find the future attractor solution by formulating a coupled dynamical set of equations. Later, we subjected the model to various data combinations and tried to understand better the influence of each cosmological data set on respective cosmological parameters.

\par
We found that while using all the datasets together, we found $H_0$ value to be in excellent agreement with the latest local determination of $H_0$ by R21 \cite{Riess:2021jrx}, though in substantial ($\sim 2.5\sigma$) tension with \textit{Planck} \cite{Planck:2018vyg} measurements. It is widely discussed that the models that alleviate the $H_0$ discrepancy do not necessarily solve other tensions, such as a higher value of $\sigma_8$ or $S_8$. Similarly, in our work, we found that the coupled model provides a remarkably better fit to the CMB, CC, BAO, Masers, PantheonPlus, $f\sigma_8$ and $H_0$ in combination with each other and predicts a (lower) late-time structure growth parameter $S_8$ = 0.757$\pm$0.023. The $S_8$ value is in good agreement with KiDS-1000-BOSS and  DES-Y3 estimations, although in moderate tension with \textit{Planck} 2018 + $\Lambda$CDM results which prefer a roughly $2.5\sigma$ higher value of $S_8$ compared to our model prediction. 
\par
We also emphasize the parameters like the sound horizon at the drag epoch $r_{drag}$ and the dark matter energy density $\Omega_{dm}$, and their respective correlations, in terms of $h$-$r_{drag}$ and $S_8$-$\Omega_{dm}$ planes. The findings predicted the smaller value of $r_{drag}$ for a larger value of $H_0$ at the cost of increasing disagreement with the \textit{Planck} data estimates. Moreover, the results from the $S_8$-$\Omega_{dm}$ parameter space showed marginal tension with \textit{Planck} with lower and constrained values of $S_8$ and $\Omega_{dm}$ for all the combined data in this work. The illustration in Fig. \ref{fig:Hz_1+z} clearly showed some disagreement ($\sim$3$\sigma$) with high-redshift BAO measurements from quasar Ly$\alpha$ observations. The interacting model confirms the overall good consistency of reconstructed $H(z)$ with combined BAO measurements of $H(z)$. Later, we also presented the time evolution of key cosmological parameters along with their confidence ranges for the coupled cosmological model. We found no significant distinction in the parameter evolution compared with the uncoupled scenario due to very low coupling parameter values preferred by the cosmological datasets.
Later, we also presented the time evolution of density parameters, the equation of state parameters, and their two-sigma confidence region for the coupled cosmological model.
\par
We then discussed the model comparison techniques to measure the goodness of fit of a given model to the data. We found that introducing extra parameters improves the fit determined by AIC and BIC and strongly favors a complex, coupled model over the $\Lambda$CDM one. Moreover, this model introduces two extra parameters, so the $``$weighted'' chi-square test favors the $\Lambda$CDM for all the combined data. Such inconsistencies and discrepancies motivate us to look further for the best theoretical model to describe the universe. Following the above analyses, these differences could signify either the need for new exotic dark energy physics to withstand the data or new observations to improve the quality of the data.  
\par
In the near future, we shall extend our analysis to incorporate one extra element which can mimic phantom dark energy at present, along with the already present interacting dark matter-dark energy scenarios to resolve present cosmological tensions.

\acknowledgments
We thank the anonymous referee for their thoughtful remarks and suggestions which led to material improvements in the paper. We are grateful to Abhijith A., Shah Navaz Adil, Anjan Ananda Sen, Elsa M. Teixeira for useful discussions. 
Part of the numerical computation of this work was carried out on the computing cluster  Pegasus of IUCAA, Pune, India.
This work is partially supported by DST (Govt. of India) Grant No. SERB/PHY/2021057. Ruchika acknowledges TASP,  iniziativa specifica INFN, for financial support. 

\appendix
\section*{Appendix} \label{Appendix}
Here our aim is to show that our results are independent of priors only when $f\sigma_8$ data is considered. To validate this, we have obtained the results by changing the ``$\sigma_8$'' range from [0.6$\sim$1] to [0.5$\sim$1], along with the updated priors (as shown in the Table \ref{table:flat priors}). The obtained results are presented in Table \ref{mean with error_sigma8_0.5-1}. Here, we have noticed the similar behaviour in the results as in the case of Table \ref{mean with error}, which shows the unconstrained $\sigma_8$ for its increased prior range, for combinations ``CMB + CC + BAO + MASERS'' and ``CMB + PantheonPlus + CC + BAO + MASERS''. Thus, indicates the prior dependence of $\sigma_8$ when $f\sigma_8$ data is excluded.

However, for other combinations that include growth rate data ($f\sigma_8(z)$), the parameter, $\sigma_8$ as well as the other key parameters are well constrained and their values are independent of the priors, as can be seen from Table \ref{mean with error} and Table \ref{mean with error_sigma8_0.5-1}. Therefore, it can be concluded that the values of the key cosmological parameters are independent of priors only when the $f\sigma_8$ is in consideration along with other data sets of CMB, SNIa, CC, BAO, MASERS, and $H_0$.
\begin{table}[!hb]
\small
\centering
\addtolength{\tabcolsep}{-4.5pt} 
\renewcommand{\arraystretch}{2}
\begin{tabular}{ c c c c c c c c c} 
\toprule [1.5pt]
  \textbf{\textit{Parameters}}  &   $\Omega_{\phi_i}*10^{-9}$   & $r_{drag}$(Mpc) & $\lambda_i$  & $m_i$ & $\Omega_{r_i}$ & $\Omega_{dm}$ & $h$ & $\sigma_{8}$\\
\bottomrule [1.5pt]
 \multicolumn{9}{c}{ CMB+CC+BAO+MASERS} \\
 \hline
 \textbf{\textit{Coupled}} & 1.125$^{+0.275}_{-0.264}$ & 147.73$^{+5.83}_{-5.86}$ & 1.03$^{+0.51}_{-0.77}$ & 0.00099$^{+0.00040}_{-0.00094}$ & 0.151$^{+0.0102}_{-0.0105}$  & 0.318$^{+0.0127}_{-0.0129}$   & 0.671$\pm$0.028 & Unconstrained  \\ 
 \bottomrule [1.3pt]
  \multicolumn{9}{c}{ CMB+PantheonPlus+CC+BAO+MASERS} \\ 
 \hline  
 \textbf{\textit{Coupled}} &  1.085$^{+0.283}_{-0.275}$ & 148.35$^{+6.06}_{-6.02}$ & 0.81$^{+0.49}_{-0.34}$ & 0.00107$^{+0.00040}_{-0.00100}$ & 0.150$^{+0.0100}_{-0.0099}$  & 0.314$\pm$0.0076   & 0.672$\pm$0.027 &  Unconstrained  \\ 
 \bottomrule [1.3pt] 
 \multicolumn{9}{c}{ CMB+PantheonPlus+CC+BAO+MASERS+$f\sigma_8(z)$} \\
 \hline
 \textbf{\textit{Coupled}} &  1.048$^{+0.284}_{-0.282}$ & 148.50$^{+5.77}_{-5.87}$ & 0.80$^{+0.49}_{-0.36}$ & 0.0012$^{+0.00047}_{-0.00110}$ & 0.151$^{+0.0095}_{-0.0096}$  & 0.314$^{+0.0076}_{-0.0077}$  & 0.671$\pm$0.026 & 0.740$^{+0.023}_{-0.022}$ \\ 
 \bottomrule [1.3pt]
 \multicolumn{9}{c}{ CMB+PantheonPlus+CC+BAO+MASERS+$f\sigma_8(z)$+$H_0$} \\
 \hline
 \textbf{\textit{Coupled}} &  1.048$^{+0.318}_{-0.307}$ & 139.11$^{+3.19}_{-3.15}$ & 0.79$\pm$0.43 & 0.0012$^{+0.00051}_{-0.00120}$ & 0.134$^{+0.0058}_{-0.0057}$  & 0.313$^{+0.0084}_{-0.0085}$   & 0.717$^{+0.015}_{-0.016}$ & 0.741$\pm$0.022 \\ 
  \bottomrule [1.3pt]
\end{tabular}
{\centering\caption{\label{mean with error_sigma8_0.5-1} Mean values with 68\% confidence level (CL) errors on cosmological and free parameters within the interacting paradigm from various data combinations with updated $\sigma_8$ [0.5$\sim$1] prior.}}
\end{table}

To solidify above inference, we also show the results with different priors (considered in Table \ref{table:different_flat priors}), on other cosmological parameters in Table \ref{appendix_meanwitherror_differentpriors}, for comparison. By comparing with Table \ref{mean with error}, we found that both the cases do not show any significant changes in key cosmological parameters such as $h$, $r_{drag}$, $\Omega_{dm}$, and $\sigma_8$ values (both mean and error bars) for ``CMB + PantheonPlus + CC + BAO + MASERS + $f\sigma_8(z)$'' and ``CMB + PantheonPlus + CC + BAO + MASERS + $f\sigma_8(z)$ + $H_0$'' combinations. Therefore, we can say that the $\sigma_8$ and other key cosmological parameters are well constrained and shows prior independence, given the growth rate data is present in the considered datasets.
\begin{table}[!ht]
\small
\centering
\addtolength{\tabcolsep}{8pt} 
\renewcommand{\arraystretch}{1.2} 
\begin{tabular}{|c | c|}
 \toprule
 \textbf{\textit{Parameters}}  & \textbf{\textit{Flat prior interval}} \\  
 \midrule
 $\Omega_{\phi_i}*10^{-9}$ & [1, 3]  \\  
 $r_{\text{drag}}$ & [130, 180] Mpc \\
 $\lambda_i$ & [$10^{-3}$, 1.5]  \\
 $m_i$ & [$10^{-5}$, $10^{-3}$] \\  
 $\Omega_{\text{r}_i}$ & [0.09, 0.17]  \\  
 $h$ & [0.5, 0.9]  \\
 $\sigma_{8}$ & [0.6, 1.0]  \\  
\bottomrule
\end{tabular}
\caption{Different priors on the parameters of the coupled model.}
\label{table:different_flat priors}
\end{table}

\begin{table}[!ht]
\small
\centering
\addtolength{\tabcolsep}{-4.5pt} 
\renewcommand{\arraystretch}{2}
\begin{tabular}{ c c c c c c c c c} 
\toprule [1.5pt]
  \textbf{\textit{Parameters}}  &  $\Omega_{\phi_i}*10^{-9}$   & $r_{drag}$(Mpc) & $\lambda_i$  & $m_i$ & $\Omega_{r_i}$ & $\Omega_{dm}$ & $h$ & $\sigma_{8}$\\
\bottomrule [1.5pt]
 \multicolumn{9}{c}{ CMB+CC+BAO+MASERS} \\
 \hline
 \textbf{\textit{Coupled}} & 1.128$^{+0.153}_{-0.149}$ & 148.21$^{+6.00}_{-6.18}$ & 0.74$^{+0.42}_{-0.42}$ & 0.00062$^{+0.00028}_{-0.00053}$ & 0.150$^{+0.0099}_{-0.0104}$  & 0.313$^{+0.0089}_{-0.0091}$   & 0.674$\pm$0.027 & Unconstrained   \\ 
 \bottomrule [1.3pt]
  \multicolumn{9}{c}{ CMB+PantheonPlus+CC+BAO+MASERS} \\
 \hline
 \textbf{\textit{Coupled}} &  1.227$^{+0.156}_{-0.151}$ & 148.45$^{+5.95}_{-6.04}$ & 0.79$^{+0.50}_{-0.29}$ & 0.00062$^{+0.00027}_{-0.00055}$ & 0.151$^{+0.0098}_{-0.0099}$  & 0.313$\pm$0.0073   & 0.672$\pm$0.027 & Unconstrained  \\  
 \bottomrule [1.3pt]
  \multicolumn{9}{c}{ CMB+PantheonPlus+CC+BAO+MASERS+$f\sigma_8(z)$} \\
 \hline
 \textbf{\textit{Coupled}} &  1.228$^{+0.156}_{-0.150}$ & 148.37$^{+5.96}_{-6.14}$ & 0.79$^{+0.49}_{-0.30}$ & 0.00061$^{+0.00026}_{-0.00055}$ & 0.151$^{+0.0100}_{-0.0102}$  & 0.314$\pm$0.0071  & 0.672$\pm$0.027 & 0.740$\pm$0.022 \\ 
 \bottomrule [1.3pt]
  \multicolumn{9}{c}{ CMB+PantheonPlus+CC+BAO+MASERS+$f\sigma_8(z)$+$H_0$} \\
 \hline
 \textbf{\textit{Coupled}} &  1.243$^{+0.169}_{-0.165}$ & 139.10$^{+3.10}_{-3.08}$ & 0.77$^{+0.54}_{-0.33}$ & 0.00066$^{+0.00032}_{-0.00059}$ & 0.134$^{+0.0057}_{-0.0056}$  & 0.312$^{+0.0079}_{-0.0078}$   & 0.717$\pm$0.015 & 0.742$\pm$0.022 \\ 
 \bottomrule [1.3pt]
\end{tabular}
{\centering\caption{\label{appendix_meanwitherror_differentpriors} Mean values with 68\% confidence level (CL) errors on cosmological and free parameters within the interacting paradigm from various data combinations using priors listed in Table \ref{table:different_flat priors}.}}
\end{table}
\newpage
\begin{center}
 \rule{4in}{0.1pt}\\
 \vspace{-13pt}\rule{3in}{0.1pt}\\
 \vspace{-13pt}\rule{2in}{0.1pt}\\
\end{center}

\begin{spacing}{0.0}
\bibliography{main}
\bibliographystyle{ieeetr}
\end{spacing} 

\end{document}